# Ultrafast Sliding Ferroelectric Switching in Bilayer Hexagonal Boron Nitride Revealed by Deep Learning Molecular Dynamics

Yinan Wang[1,2], Poyen Chen[1,2], Teruyasu Mizoguchi[2]

1. Department of Materials Engineering, the University of Tokyo, Tokyo, Japan
2. Institute of Industrial Science, the University of Tokyo, Tokyo, Japan

**Abstract**

Sliding ferroelectricity in bilayer hexagonal boron nitride (h-BN) offers compelling prospects for next-generation non-volatile memory, yet the atomistic dynamics of electric-field-driven polarization switching remain poorly understood. Here, we present a fully data-driven, coupled atomistic framework that integrates a fine-tuned MACE machine learning potential (MLP) with an equivariant graph convolutional neural network (EGCNN) for real-time Born effective charge (BEC) prediction, enabling large-scale non-equilibrium molecular dynamics simulations of AB-stacked bilayer h-BN under applied electric fields. By implementing a rigorous real-space path-integral polarization formalism combined with a state-constrained Gaussian convolution background extraction procedure, we successfully isolate the intrinsic spontaneous polarization from the dominant dielectric background. Our simulations reveal that coherent single-domain rigid sliding, completing within 5 ps, constitutes a physically viable ultrafast switching mechanism, and reproduces clean ferroelectric hysteresis loops whose shape is qualitatively consistent with experimental observations.

**Introduction**

Conventional ferroelectrics rely on the cooperative displacement of ions within a polar crystal lattice to generate and reverse spontaneous polarization. In two-dimensional (2D) van der Waals (vdW) materials, a fundamentally distinct mechanism has emerged, sliding ferroelectricity, in which macroscopic interlayer translation between specific stacking configurations breaks inversion symmetry and produces a switchable out-of-plane electric polarization.[1–4] First predicted from first-principles calculations by Li and Wu, this phenomenon was experimentally confirmed in 2021 through the simultaneous discovery of stacking-engineered ferroelectricity in parallel-stacked bilayer h-BN and interfacial ferroelectricity driven by vdW sliding.[1,2,4] These landmark observations established a new paradigm, termed slidetronics, in which the stacking degree of freedom, rather than the soft phonon mode, serves as the primary order parameter governing the ferroelectric state.[2,5]

Among the growing family of 2D sliding ferroelectrics, which now encompasses rhombohedral transition metal dichalcogenides (TMDs), γ-InSe, and even multilayer graphene systems, bilayer h-BN occupies a singular position as the cleanest and most technologically compelling prototype.[6–9] Its wide bandgap, minimal leakage current, and dual role as both ferroelectric active layer and gate dielectric render it uniquely suited for device integration. The intrinsic out-of-plane spontaneous polarization of AB-stacked bilayer h-BN is modest, yet this is sufficient to produce pronounced electrostatic effects at the monolayer limit. Crucially, the interlayer sliding barrier is exceptionally shallow, enabling switching at remarkably low coercive fields.[4] The technological promise of this material has been dramatically validated by recent demonstrations of ferroelectric field-effect transistors (FeFETs) based on bilayer h-BN, which achieve nanosecond-scale switching speeds and endurance exceeding $10^{11}$ cycles, figures rivaling state-of-the-art $HfO_2$-based memories.[10] Concurrently, sliding ferroelectric $3R-MoS_2$ devices have exhibited fatigue-free operation over $10^6$ cycles without wake-up degradation, and epitaxial h-BN grown on graphene has demonstrated robust ferroelectric switching with remanent polarization of ~0.375 μC/cm², collectively establishing sliding ferroelectrics as serious contenders for next-generation non-volatile memory and neuromorphic computing.[11–14]

Beyond the homogeneous bilayer, marginally twisted h-BN hosts a rich landscape of emergent polar phenomena. Small twist angles produce moiré superlattices with periodic triangular networks of charge-polarized interfacial domains, within which the continuously varying local stacking configuration gives rise to topologically non-trivial polar textures.[3] Bennett et al. demonstrated theoretically that these moiré domains support polar meron–antimeron networks, where the combined in-plane and out-of-plane polarization vector winds with half-integer topological charge.[15] Operando electron microscopy has directly visualized domain wall (DW) dynamics in twisted vdW homobilayers, revealing domain wall velocities of ~300 μm/s and pronounced Barkhausen-like pinning effects under gate-voltage cycling.[16] More recently, vector piezoresponse force microscopy has confirmed the existence of Bloch-type meron/antimeron networks in marginally twisted h-BN, and electric-field manipulation of individual moiré ferroelectric domains has been achieved.[17,18] These findings underscore that the switching kinetics, domain nucleation pathways, wall mobility, topological transitions, and thermal activation, are intrinsically dynamical phenomena that cannot be captured by static calculations alone.

Despite the urgent need for dynamical insight, existing theoretical approaches face fundamental limitations. Standard density functional theory (DFT) and density functional perturbation theory (DFPT) provide accurate equilibrium polarization and energy barriers, but introducing finite electric fields under periodic boundary conditions leads to well-documented convergence difficulties, precluding direct ab initio molecular dynamics (AIMD) simulation of field-driven switching at experimentally relevant length and time scales. [4,15,19–21] Recent MLP studies have made significant strides. He et al. employed an MLP to simulate domain wall dynamics in bilayer h-BN, resolving a ~10 nm wall width and demonstrating that domain wall motion reduces the critical switching field by approximately two orders of magnitude relative to homogeneous nucleation.[22] However, their polarization was derived from a phenomenological mapping to the Berry phase at selected commensurate stackings, an approximation that breaks down for non-commensurate moiré textures, in-plane polarization components at domain walls, and the strongly environment-dependent BECs encountered during dynamic sliding. More recently, Deng et al. demonstrated a multi-task equivariant neural network (DREAM-Allegro) that simultaneously predicts interatomic forces and BECs for bilayer h-BN, and Falletta et al. developed a unified differentiable framework enforcing exact physical constraints on electric enthalpy and polarization for bulk ferroelectrics.[21,23] Nevertheless, neither approach has delivered fully coupled, real-time structural and polarization dynamics for 2D sliding ferroelectrics under time-varying electric fields, the regime most directly relevant to device operation.

To bridge this critical gap, we present a fully data-driven atomistic simulation framework that couples a fine-tuned MACE machine learning potential with an EGCNN for real-time BEC prediction, enabling non-equilibrium molecular dynamics of bilayer h-BN under applied electric fields with simultaneous, first-principles-quality tracking of the polarization evolution via a rigorous real-space path-integral formalism.[24,25] Our simulations reveal that coherent single-domain rigid sliding, completing within 5 ps, constitutes a physically viable ultrafast switching mechanism, and reproduces clean ferroelectric hysteresis loops consistent with experimental observations. This framework establishes a computational foundation for the predictive atomistic design of 2D sliding ferroelectric devices.

**Results & Discussion**

**Structural Dynamics and Thermodynamic Stability**

One of the critical challenges in developing MLP for layered 2D materials is the accurate description of vdW interactions. A prevalent computational strategy is to involve linearly superimposing an empirical dispersion correction, such as DFT-D3BJ, onto the baseline MLP. Mikkel *et al.* have extensively discussed this method.[26,27] When the D3BJ correction was explicitly coupled with the MACE-MP-0 model, the structural relaxation of the thermodynamically stable AA′ stacking yielded an expanded interlayer distance of 3.85 Å. This value overestimates the experimentally established interlayer spacing of approximately 3.30 Å.[28,29] Consequently, we ensured the intrinsic capture of the vdW interactions by strictly fine-tuning the MACE model on an *ab initio* dataset computed with the non-local vdW-DF2-B86R exchange-correlation functional.[30] The AA′ stacking interlayer distance computed by both the vdW-DF2-B86R functional (DFT) and the fine-tuned MACE model is approximately 3.29 Å, exhibiting quantitative agreement with experimental measurements.

The fine-tuned MACE model exhibits robust predictive capabilities for both potential energies and atomic forces across a diverse structural phase space, encompassing monolayer, various bilayer stackings and complex moiré superlattices. Quantitative evaluation yields a training root-mean-square error (RMSE) of 19.3 meV/atom for energies and 62.3 meV/Å for forces. This exceptional parity between training and validation metrics indicates a well-regularized potential free from overfitting, which is further corroborated by the tightly grouped linear correlations in the energy and force parity plots across all data subsets.

Beyond static energetic accuracy, we assessed the harmonic lattice dynamical accuracy of the fine-tuned MACE potential. We verified this by computing the phonon dispersion relations for both the AA′ and AB bilayer configurations. The ML-predicted phonon band structures display excellent agreement with rigorous DFT benchmark calculations along the high-symmetry Brillouin zone path. The accurate reproduction of both acoustic and optical phonon branches confirms that the fine-tuned MACE model properly describes the harmonic interatomic force constants and accurately captures the underlying vibrational thermodynamics.

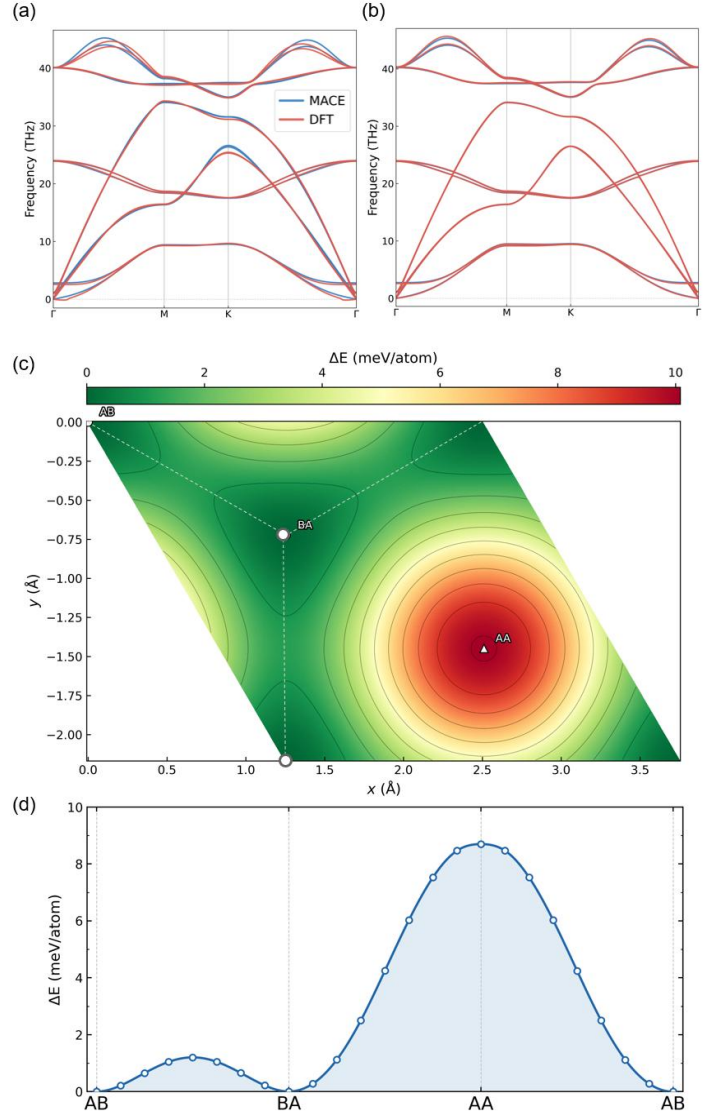

Phonon dispersion relations and interlayer sliding energetics of bilayer h-BN. (a, b) Phonon dispersion relations along the Γ–M–K–Γ high-symmetry path for the (a) AA′ and (b) AB stacking configurations, calculated using the fine-tuned MACE machine learning potential (blue lines) and DFT with the vdW-DF2-B86R functional (red lines). (c) Two-dimensional potential energy surface (PES) for rigid interlayer sliding of the AB-stacked bilayer, computed with the fine-tuned MACE potential. The AB and BA stackings correspond to degenerate global energy minima, while the AA configuration constitutes the global energy maximum. The color scale indicates relative energy ΔE in meV/atom. (d) Minimum energy pathway (MEP) along the AB→BA→AA→AB trajectory from CI-NEB calculations. The switching pathway proceeds via a low-barrier saddle point, bypassing the high-energy AA configuration and yielding an exceptionally low kinetic barrier consistent with facile coercive-field switching.

To elucidate the physical mechanism underlying the sliding ferroelectric transition, we first mapped the potential energy surface (PES) of bilayer h-BN using the fine-tuned MACE model (Fig.1c). The AB and BA stacking configurations emerge as degenerate global energy minima. Conversely, the fully eclipsed AA stacking configuration corresponds to the global energy maximum, with a substantial energy penalty of approximately 8.70 meV/atom relative to the AB/BA stackings.

Nudged elastic band (NEB) calculations reveal that the minimum energy pathway (MEP) for the stacking inversion strictly follows the [$1\bar{1}0$] crystallographic direction or its symmetry equivalents (Fig.1d). Crucially, when transitioning from the AB to the BA state, the system avoids the high-energy AA state entirely. Instead, the sliding pathway proceeds through an intermediate saddle point (SP), following the trajectory: AB→SP→BA→SP→AB. The kinetic energy barrier at this saddle point is exceptionally low. This highly shallow energy landscape facilitates the experimentally observed low coercive fields for polarization switching, but it simultaneously implies that the system is highly susceptible to spontaneous, thermally driven domain sliding.

**Born Effective Charge and Polarization Prediction**

To predict the real-time BEC tensors under varying structural configurations, an EGCNN was trained. Our previous studies commonly utilized a fine-tuning approach on pre-trained BM1 model, which typically employ a local graph cutoff radius of 3.0 Å.[25,31–34] While a 3.0 Å cutoff is generally sufficient for predicting the properties of covalently bonded bulk materials, it inherently fails for layered van der Waals materials like h-BN. Because the nominal interlayer distance of bilayer h-BN strictly exceeds this predefined cutoff, the message-passing mechanism within such models cannot learn the crucial out-of-plane interlayer electrostatic interactions. To resolve this spatial limitation, we adopted the exact EGCNN architecture as the BM1 model but expanded the local cutoff radius to 4.5 Å.

Evaluating the trained EGCNN model on an independent inference set containing 45,008 atoms reveals excellent local precision across all structural configurations (Fig.2a). Remarkably, the errors for the intermediate sliding pathways and the monolayer structures fall to an exceptionally low magnitude of $2.0 \times 10^{-4}\ e^-$, while the complex moiré superlattices maintain a low MAE of $8.0 \times 10^{-3}\ e^-$. To contextualize this charge accuracy for atomistic simulations, we evaluate the resulting force deviation under applied external fields. In our sliding ferroelectric MD simulations, the maximum applied out-of-plane electric field is 2.0 V/Å. Given a conservative maximum BEC error, the maximum corresponding field-induced force error is strictly less than 17 meV/Å. This deviation is substantially smaller than the intrinsic force RMSE of the fine-tuned MACE potential, confirming that the EGCNN model provides highly reliable local electrostatic coupling without acting as the accuracy bottleneck.

However, it is critically important to note that while the EGCNN achieves exceptional local predictive precision, the independently generated machine-learned tensorial outputs do not inherently satisfy global physical conservation laws. Specifically, the raw predicted BEC marginally deviate from the acoustic sum rule (ASR). Uncorrected, even negligible residual pseudo-charges can accumulate over large simulation cells and artificially skew the macroscopic polarization curve during continuous dynamic sliding. To rigorously enforce global charge neutrality and maintain physical fidelity, an ASR correction is systematically applied to the raw EGCNN predictions at every integration step before evaluating the field-induced forces and tracking the polarization evolution.

To visually demonstrate the critical necessity of this ASR correction and to further validate the predictive fidelity of our EGCNN model, we calculated the continuous out-of-plane polarization $P_z$ along the minimal energy sliding pathway using the path-integral formulation (Fig.2b). As depicted by the red dashed line, directly integrating the raw, uncorrected BECs leads to a pronounced, unphysical macroscopic drift.

In contrast, integrating the ASR-corrected effective charges completely eliminates this accumulated artifact. The resulting trajectory (solid blue line) forms a closed, continuous sinusoidal curve that correctly reflects the periodic structural symmetry of the sliding pathway. More importantly, this ASR-corrected continuous polarization profile exhibits agreement with the rigorous *ab initio* Berry phase calculations and effective charge evaluations reported in previous theoretical literature.[6,15,22,35] The successful reproduction of this intrinsic polarization curve serves as compelling evidence that our EGCNN model not only achieves high local accuracy but also correctly captures the complex, non-linear dielectric dynamics of the atomic layers during continuous sliding, providing a fully reliable electronic foundation for the subsequent large-scale dynamic simulations.

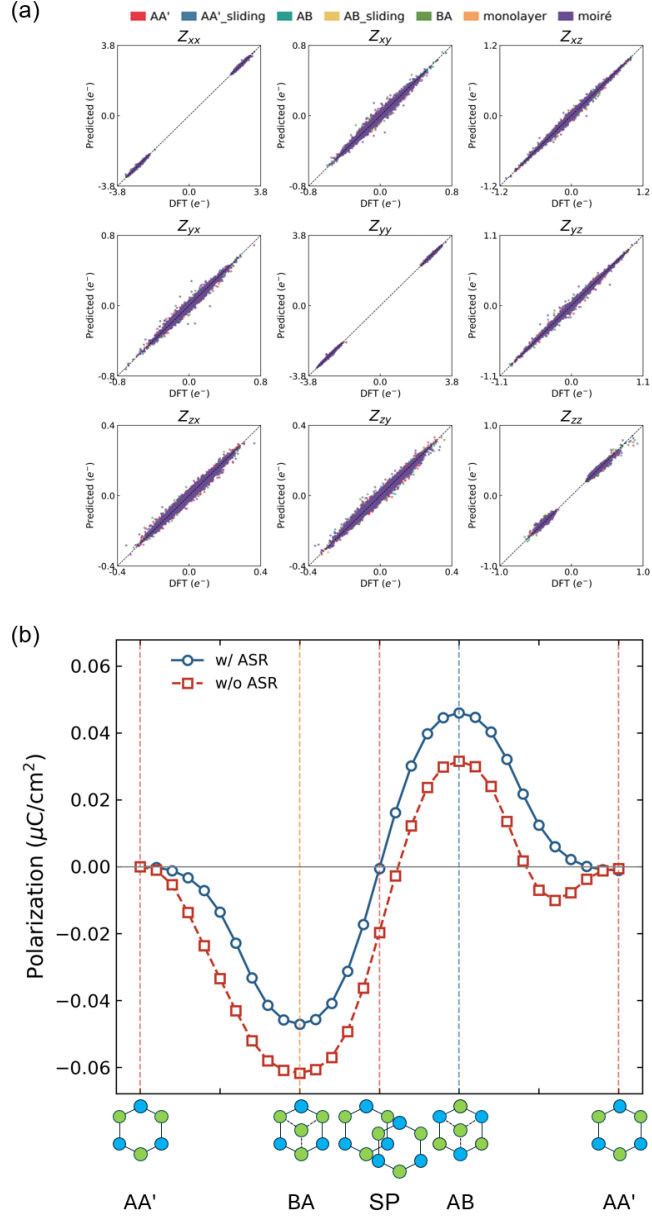

Fig.2 Performance of the EGCNN Born effective charge model and validation of the path-integral polarization formalism. (a) Parity plots of all nine BEC tensor components ($Z^*_{i,\alpha\beta}$) comparing EGCNN predictions against DFT values on an independent inference set of 45,008 atoms, color-coded by structural subset: AA′, AA′ sliding, AB, AB sliding, BA, monolayer, and moiré superlattices. The tight linear correlations across all components and structural phases confirm the high local predictive fidelity of the EGCNN model. (b) Continuous out-of-plane polarization $P_z$ calculated along the minimum energy sliding pathway using the real-space path-integral formulation, with and without ASR correction applied to the EGCNN-predicted BECs.

**E-Field Driven Dynamics**

To investigate the dynamic polarization switching behavior of bilayer h-BN under realistic external electric fields, we performed non-equilibrium MD simulations. Within our coupled machine-learning framework, the total force acting on each atom at every integration step is dictated by the sum of the interatomic force derived from the MACE potential and the electrostatic force induced by the external field, as detailed in the Methods section. The simulations were initially carried out on a 20 Å supercell at 50 K, applying a cyclic, step-wise time-varying out-of-plane electric field.

The temporal structural evolution, parameterized by the AB stacking state alongside the applied electric field, is presented in Fig.3a. As the external field progressively ramps to a maximum amplitude of $\pm 2.0$ V/Å, the system exhibits sharp, deterministic structural phase transitions. Specifically, the stacking configuration abruptly switches between AB and BA precisely when the applied field reaches the critical coercive threshold.

Detailed trajectory analysis reveals that the complete structural inversion, corresponding to a macroscopic interlayer translation of one B-N bond length, is accomplished within an ultrafast time window. To directly visualize the atomistic mechanism of this switching event, representative snapshots extracted from the MD trajectory are presented in Fig.3c–d. At 162 ps, the bilayer retains the AB stacking configuration immediately prior to switching. At 163 ps, the structure transiently occupies the saddle-point geometry along the minimum energy pathway, wherein neither the AB nor the BA registry is established. By 164 ps, the transition is complete and the system has adopted the BA stacking state. The reverse BA→AB transition, captured at 563–565 ps (Fig.3d), proceeds analogously via the symmetry-equivalent SP stacking. To structurally quantify the switching kinetics, the RMSD of atomic positions was computed at the two coercive field steps (Fig.3e,f). At both $\varepsilon_z = -1.6$ V/Å and $+1.6$ V/Å, the RMSD exhibits a sharp transient spike coinciding with the stacking inversion, followed by convergence to a new stable plateau. Inspection of the atomic trajectories confirms that the stacking inversion itself is complete within approximately 5 ps, while full thermodynamic relaxation to the new equilibrium registry requires approximately 7 ps. Specifically, the switching duration is approximately 5 ps at 50 K and decreases to approximately 2 ps at 300 K, reflecting thermal-fluctuation-assisted acceleration of the sliding dynamics at elevated temperatures. Each switching event spans approximately 2,000~5,000 MD integration steps of 1 fs, providing sufficient temporal resolution to fully resolve the transition dynamics. Crucially, both the 2 ps and 5 ps switching timescales are substantially longer than the characteristic period of h-BN optical phonons, which are up to 20~25 fs, corresponding to the 40~50 THz frequencies observed in our calculated phonon dispersion, Fig.1. This confirms that the atomic motion during switching is fully thermalized and not ballistic. In this sense, the simulated switching timescales carry well-defined physical meaning, even though a direct quantitative mapping to macroscopic experimental switching times, which are governed by domain nucleation and wall propagation at much longer length scales, is not implied.

Inspection of the dynamic trajectory further demonstrates that this structural transition involves the simultaneous, concerted translation of both the upper and lower layers along the $[\bar{1}\bar{2}0]$ crystallographic direction. Interestingly, the forward switching from AB to BA and the subsequent backward reversal from BA to AB do not retrace the same spatial route. Instead, they proceed along distinct $[0\bar{1}0]$, symmetrically equivalent to $[1\bar{1}0]$. This structural hysteresis is in agreement with the minimum energy pathways mapped in our static PES and NEB calculations (Fig.1). While previous experimental and theoretical studies have predominantly reported polarization switching governed by localized domain nucleation and subsequent DW propagation, our simulations provide compelling theoretical evidence that coherent single-domain rigid sliding is also a physically viable switching mechanism under sufficient field strengths, enabling exceptionally fast reversal dynamics in defect-free lattices. [1,3,10,15,22,36]

The simulated $\varepsilon_c$ ($\pm 1.5$ V/Å) is distinct from a naive thermodynamic estimate, which balances the electric enthalpy against the switching barrier using only the $P_z$. As detailed in the SI, such a simplistic electrostatic projection yields an unphysical intrinsic coercive field of approximately 39.8 V/Å. This overestimation demonstrates that the switching mechanism is not solely governed by the direct coupling between the $\varepsilon_z$ and $P_z$. Instead, the dominant driving force originates from the off-diagonal BEC components ($Z^*_{zx}$ and $Z^*_{zy}$). These tensors efficiently transduce the applied out-of-plane electric field into in-plane atomic forces along the sliding direction. This electromechanical transduction, naturally captured by our fully coupled MACE-EGCNN framework, provides a rigorous microscopic rationalization for how realistic field strengths can drive macroscopic interlayer sliding despite the marginal magnitude of $P_z$.

It is noteworthy that the critical coercive field $\varepsilon_c$ required to drive this single-domain switching in our simulations is substantially higher than bilayer h-BN experimental values.[1,12] A recent unified machine-learning study on BaTiO$_3$ similarly founds that the simulated ideal single-crystal coercive field exceeds experimental values by approximately an order of magnitude due to the absence of pre-existing domains and interfacial defects.[23] This is very similar to our situation. As experiments observe the formation and movement of DW, while our MD observes intrinsic rigid translations.[10,12,37,38] Therefore, the substantially higher coercive field in our simulations is a direct physical consequence of overcoming the macroscopic kinetic barrier all at once, without the aid of localized DW motion.

To elucidate the physical origin of the $\varepsilon_c$ further, we evaluated the coercive field across various supercell dimensions and operating temperatures. The critical field shows only a marginal thermal reduction, decreasing slightly from 1.5 V/Å at 50 K and 100 K to 1.4 V/Å at 200 K and 300 K. (Fig.S5) Furthermore, we observe that different random initial velocity seeds introduce a slight variation in the $\varepsilon_c$. Taking this stochastic uncertainty into account, we conclude that increasing temperature does not produce a statistically significant change in the coercive field. This near-temperature-independence of $\varepsilon_c$ stands in marked contrast to conventional displacive ferroelectrics like BaTiO$_3$. In BaTiO$_3$, the $\varepsilon_c$ decreases significantly with increasing temperature, which is a hallmark of thermally activated domain nucleation where thermal fluctuations progressively lower the switching barrier. [39] Conversely, the coercive field in our h-BN simulations remains remarkably temperature-independent. This stark contrast highlights a qualitatively distinct switching physics. In sliding h-BN, the transition is dominated by the aforementioned electromechanical in-plane force mechanically overcoming a shallow vdW potential landscape. Consequently, the switching dynamics are fundamentally insensitive to thermal activation, an athermal characteristic that explicitly sets 2D sliding ferroelectricity apart from conventional displacive mechanisms.

Concurrently, the macroscopic $P_z$ along the dynamic trajectory was continuously tracked using the real-space path-integral formalism over the predicted BECs (Fig.3b). While the raw $P_z$ curve captures the primary polarization reversals associated with the structural sliding, the polarization evolution is not strictly synchronized with the stacking transitions. Specifically, at the high-field boundaries where the stacking inversion has already completed, $P_z$ continues to exhibit a linear dependence on the magnitude of the applied electric field. Consequently, the dynamic polarization trajectory diverges from an ideal flat hysteresis behavior, displaying four distinct and symmetric segments characterized by field-dependent linear slopes. Furthermore, due to the continuous cumulative nature of the path integral, the maximum absolute magnitude of the integrated polarization clearly exceeds the intrinsic spontaneous polarization predicted by static zero-field calculations (Fig.2b). These observations indicate that the raw polarization signal is a superposition of the spontaneous sliding polarization and an accumulating field-dependent background. This physical feature underscores the necessity of a systematic background extraction procedure.

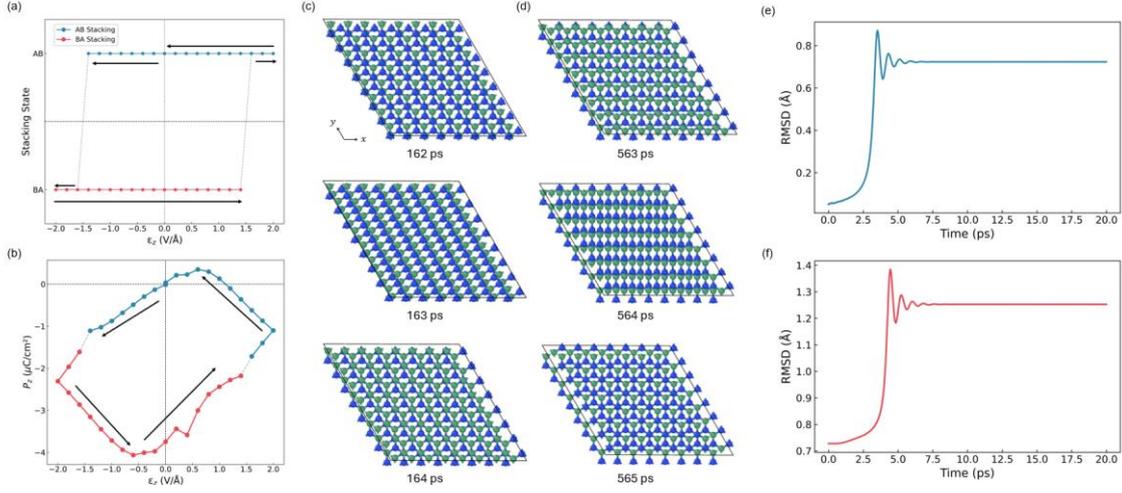

Fig.3 Electric-field-driven structural switching dynamics, atomic trajectory snapshots, and root-mean-square displacement (RMSD) analysis of bilayer h-BN. (a) Temporal evolution of the stacking configuration (AB or BA) as a function of the applied out-of-plane electric field ε$z$ (V/Å) during a full cyclic sweep (0→−2→0→+2→0 V/Å) at 50 K in a 20 Å supercell. (b) Corresponding raw cumulative $P$z as a function of the applied field. (c) Representative atomic trajectory snapshots at 162, 163, and 164 ps viewed along the out-of-plane direction, capturing the AB→BA transition. The structure at 162 ps retains the AB stacking configuration; the snapshot at 163 ps corresponds to the transient SP geometry; and the structure at 164 ps has completed the transition into the BA stacking state. (d) Analogous snapshots at 563, 564, and 565 ps illustrating the reverse BA→AB transition. Green and blue spheres represent boron and nitrogen atoms, respectively. (e) Time evolution of the total atomic RMSD computed relative to the initial AB-stacked reference configuration (t = 0) at $\varepsilon_z$ = −1.6 V/Å, corresponding to the AB→BA switching event. (f) Equivalent RMSD analysis at $\varepsilon_z$ = +1.6 V/Å for the BA→AB transition.

For a static, zero-field h-BN (Fig.4a), the macroscopic polarization originates intrinsically from the spontaneous symmetry breaking induced by the specific stacking configuration.[6] However, upon the application of an external electric field (Fig.4b), the oppositely charged boron and nitrogen atoms experience divergent out-of-plane electrostatic forces. To re-establish mechanical equilibrium, the MACE potential provides an interatomic restoring force to counterbalance this electrostatic pull, effectively acting as a harmonic spring model. Consequently, an out-of-plane elastic displacement proportional to the external field is induced. According to our path-integral formulation (Eq.1), this field-induced atomic displacement yields an additional out-of-plane polarization $\Delta P_z$ that scales linearly with the applied field $\varepsilon_z$. Due to the continuous nature of the path integral, this generates a massive macroscopic dielectric background proportional to the cumulative field $\sum_{t=0}^{T} \varepsilon_z(t)$, which is fundamentally independent of the stacking orientation. Compounded by complex intra- and inter-layer electromechanical balances, alongside superimposed high-frequency thermal vibrations, this cumulative background ultimately dwarfs the intrinsic spontaneous polarization by a significant margin. As evidenced in Fig.3b, the true ferroelectric signal is entirely swamped by this dielectric response, rendering simple mathematical fitting practically ineffective for background removal. Moreover, as the field-induced displacements are intrinsically embedded in the displacement record itself, decoupling this dielectric contribution via displacement-component decomposition proves equally intractable.

Direct atomistic evidence for this field-induced displacement mechanism is provided by tracking the mean out-of-plane displacement $\langle \Delta z \rangle$ of the boron and nitrogen sublattices separately throughout the full cyclic sweep (Fig.4e). Under zero field, both sublattices remain at their equilibrium out-of-plane registry. As the applied field ramps up, the boron and nitrogen atoms develop oppositely directed $\langle \Delta z \rangle$ of progressively increasing magnitude, consistent with the divergent electrostatic forces $F_{ex} = |e| \varepsilon_z Z^*_{i,\alpha z}$ acting on the two sublattices as described in Fig.4b. This field-proportional, antisymmetric sublattice breathing confirms that the dominant contribution to the cumulative path-integral polarization is of elastic, dielectric origin rather than ferroelectric. Furthermore, it is noteworthy that the field-proportional growth of $\langle \Delta z \rangle$ is entirely independent of the instantaneous stacking configuration. Even across the switching events at approximately 160 ps and 560 ps, where the stacking state abruptly inverts between AB and BA, the magnitude of the sublattice displacement continues to increase monotonically with the applied field without interruption. This stacking-independence directly demonstrates that the dielectric background is governed exclusively by the electromechanical response of the B–N bond to the applied field, and carries no information about the ferroelectric order parameter. It further validates the physical premise of the state-constrained background extraction procedure: because the dielectric background evolves continuously and independently within each stacking state, segmenting the trajectory by stacking configuration and applying the Gaussian convolution filter separately to each segment correctly isolates the field-induced contribution without distorting the intrinsic spontaneous polarization.

To disentangle the intrinsic ferroelectric signal from this complex temporal trajectory, we exploited the distinct frequency characteristics of the superposed components. The external-field-induced dielectric background acts as a low-frequency, time-dependent signal, whereas the thermal atomic motion manifests as high-frequency, time-independent noise. To isolate these components without artificially smearing the intrinsic polarization jump during the structural phase transition, we implemented a state-constrained Gaussian convolution procedure serving as a piecewise low-pass filter. Specifically, the temporal trajectory was partitioned based on the instantaneous stacking configuration, and the convolution was applied separately to the continuous AB and BA segments. This mathematically smooths out the high-frequency thermal fluctuations within each distinct structural phase, extracting the pure field-induced dielectric polarization. The mathematical formulations are extensively detailed in the Methods section. As illustrated in Fig.4d, because the filtering is performed independently for each stacking state, the convoluted background faithfully tracks the field-dependent global trend of the raw signal while preserving the abrupt discontinuous steps precisely synchronized with the structural stacking inversions. By subtracting this extracted background from the raw $P_z$ trajectory, we successfully unmasked the true polarization evolution, recovering the intrinsic spontaneous switching alongside the natural thermal dynamics. The resulting purified data yields the remarkably clean and flat intrinsic hysteresis loop presented in Fig.4c.

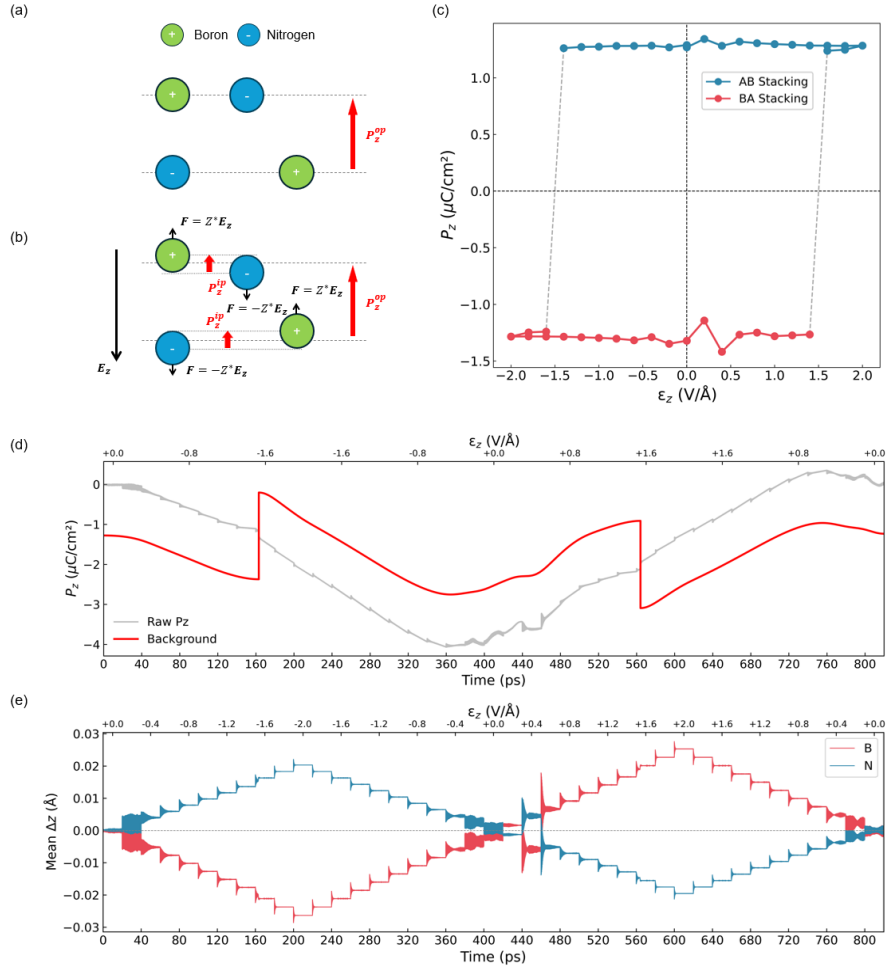

Fig.4 Physical origin of the dielectric background and extraction of the intrinsic ferroelectric hysteresis loop. (a) Schematic of the spontaneous $P_z^{op}$ in zero-field AB-stacked bilayer h-BN arising from interfacial symmetry breaking, with boron (positive, green) and nitrogen (negative, blue) sublattices displaced vertically. (b) Schematic illustrating field-induced out-of-plane elastic displacements under an applied field $\varepsilon_z$. Electrostatic forces $F_{ex} =|e|\varepsilon_z Z^*_{i,\alpha z}$ act divergently on the oppositely charged B and N sublattices, generating field-proportional intra- and inter-layer displacements ($P_z^{ip}$) that superimpose on the spontaneous polarization $P_z^{op}$, producing a cumulative dielectric background via the path-integral formulation. (c) Purified ferroelectric hysteresis loop obtained by subtracting the extracted background from the drift-corrected $P_z$ trajectory. (d) Raw cumulative $P_z$ trajectory (gray) as a function of time alongside the extracted state-constrained Gaussian convolution background (red). (e) Time evolution of the mean out-of-plane displacement ⟨Δz⟩ of boron (red) and nitrogen (blue) atoms throughout the full cyclic field sweep.

The purified data, presented in Fig.4d, yields a highly symmetric and well-defined ferroelectric hysteresis loop. The shape of the hysteresis loop is consistent with prior experimental observations, with the polarization remaining nearly invariant within the same stacking state.[1] In stark contrast to the pronounced linear slope observed in the raw curve (Fig.3b), the purified loop exhibits remarkably flat saturation plateaus in the high-field regime. Upon structural switching, the polarization magnitude undergoes an abrupt jump and subsequently stabilizes at a constant value dictated entirely by the newly adopted stacking state. This saturation contrasts with the continuous, field-dependent polarization increase seen in Fig.3b. The successful reproduction of these hallmark features serves as the definitive criterion validating our background subtraction methodology. It unequivocally demonstrates that the field-induced dielectric background has been precisely decoupled from the total signal, thereby highlighting the true spontaneous polarization jump at the exact instant of structural inversion.

While the extracted remanent polarization successfully reveals the intrinsic bistability of the sliding ferroelectric states, its absolute magnitude remains notably higher than values derived from prior static calculations and existing experimental measurements.[1,22] This discrepancy implies that during the non-equilibrium dynamic simulation, certain complex contributions tied to atomic dynamics, such as persistent thermal fluctuations, have not been entirely isolated. Constrained by current signal processing capabilities, a more refined decoupling strategy to further minimize this quantitative variance is presently lacking. This highlights the inherent complexity of addressing cumulative dynamic polarization effects within a continuous path-integral framework.

Finally, it is necessary to acknowledge the operational limitations of the current background extraction scheme. The methodology relies heavily on the predetermined classification of the system's stacking state (AB or BA). While this piecewise processing logic excels at resolving coherent, single-domain sliding, it faces significant challenges when applied to structurally complex systems. For instance, in moiré-stacked h-BN superlattices, which feature continuous displacement gradients and localized spatial polarization distributions, the local stacking configuration evolves continuously across the spatial domain.[15,21,35] Consequently, extending this state-discrimination-based algorithm to such systems will necessitate the development of more generalized and sophisticated mathematical models.

**Methods**

**Structural Modeling and Data Preparation**

The initial atomic coordinates of the hexagonal boron nitride (h-BN) primitive cell were obtained from the Materials Project database (mp-984).[40,41] Based on this unit cell, we constructed monolayer h-BN alongside various bilayer configurations, including AA′, AB, and BA stackings. These fundamental structures were expanded into $4 \times 4 \times 1$ supercells, yielding 64 atoms for the bilayer structures. To map the potential energy surface and generate diverse training data, the top layer of the AA′ and AB configurations was rigidly translated along the $[1\bar{1}0]$ crystallographic direction, covering a full period of stacking transitions. This continuous sliding process generated 300 distinct configurations for both the AA′ and AB sliding pathways. Additionally, moiré superlattices, specifically $\sum 7$, $\sum 13$, $\sum 19$ commensurabilities, were generated using the InterfaceMaster Python package.[42,43] To eliminate interactions between periodic images along the out-of-plane direction, a 15 Å vacuum layer was introduced for all simulated structures using the VASPKIT code.[44]

**First-Principles Calculations**

All first-principles calculations were carried out within the framework of density functional theory (DFT) as implemented in the Vienna Ab initio Simulation Package (VASP). The Brillouin zone was sampled using a Γ-centered k-point mesh with a uniform density of 0.025 $Å^{-1}$ in the xy-plane. The plane-wave energy cutoff was set to 750 eV.

Structural relaxations, ab initio molecular dynamics (AIMD) with on-the-fly machine learning force fields, and phonon dispersion calculations were performed utilizing the non-local vdW-DF2-B86R exchange-correlation functional. The reason why this functional was specifically chosen is discussed in the Results section. The atomic positions were fully relaxed until the energy convergence threshold reached 0.01 eV.

To sample the finite-temperature configurational space, AIMD simulations were performed for the AA′, AB, and BA stackings. These simulations were carried out in the canonical (NVT) ensemble using a Nosé-Hoover thermostat, spanning a temperature range from 1 K to 2500 K. Each trajectory was evolved for 30,000 steps, from which uncorrelated snapshots along with their corresponding total energies and atomic forces were extracted for subsequent model training. The phonon dispersion relations were calculated utilizing the finite displacement method as implemented in the Phonopy code.[45]

Finally, the Born effective charge (BEC) tensors were computed via density functional perturbation theory (DFPT). For these specific charge calculations, we employed the generalized gradient approximation (GGA) parameterized by the Perdew-Burke-Ernzerhof (PBE) functional, in conjunction with the projector augmented-wave (PAW) method.[46–48] The electronic self-consistent loop was converged to a strict threshold of $1 \times 10^{-5}$ eV.

**Machine Learning Potential Fine-Tuning**

To accurately capture the interatomic forces and potential energy surface during the MD simulation, we fine-tuned the pre-trained MACE-MP-0 model.[24] The training dataset comprised a total of 2808 structures, incorporating both configurations sampled from AIMD trajectories and relaxed sliding structures. To ensure a balanced representation of the configurational space and prevent dataset bias, the structural data was subjected to a stratified split based on the specific stacking types. For each individual stacking category, the data was partitioned into a training set and a testing set maintaining a strict 8:2 ratio. During the model training phase, 10% of the training subset was randomly allocated as a validation set to monitor convergence and prevent overfitting.

**Equivariant Graph Convolutional Neural Network for BEC Prediction**

The EGCNN model was instead trained entirely from scratch. The training corpus consisted of 4045 DFT-computed structures with rigorously evaluated BEC tensors. Consistent with the MACE training protocol, we applied a stratified 8:2 split for the training and testing sets across all individual structural configurations to ensure robust generalizability of the predicted dielectric properties.

**Molecular Dynamics under Applied External Electric Fields**

All MD simulations were performed using the Atomic Simulation Environment (ASE) Python library.[49] To simulate the structural evolution under external electric fields, we dynamically coupled the structural and electronic predictions. At each integration step, the interatomic forces ($F_{MLP}$) were evaluated using the fine-tuned MACE potential. Simultaneously, the instantaneous BEC tensor for each atom in the current configuration was predicted using the EGCNN model. The application of an external electric field $\varepsilon_\beta$ introduces an additional electrostatic force ($F_{ex}$) on each atom, defined as $F_{ex} = |e|\varepsilon_\beta Z^*_{i,\alpha\beta}$, where $|e|$ is the elementary charge. Consequently, the total force governing the atomic trajectory at any given step is precisely given by $F_{total} = F_{MLP} + F_{ex}$.

For the electric-field-driven sliding simulations, we selected a 20.08 Å supercell (comprising 256 atoms). These simulations were similarly conducted in the NVT ensemble with a Nosé-Hoover thermostat at specific temperatures of 50 K, 100 K, 200 K, and 300 K. To capture the full ferroelectric hysteresis loop, a cyclic, stepwise out-of-plane electric field was applied along the sequence of 0→-2→0→2→0 V/Å, utilizing a uniform sweep increment of 0.4 V/Å. To guarantee comprehensive structural relaxation and thermal equilibration at each field intensity, the system had evolved for a duration of 20 ps per field step, integrating the equations of motion with a 1.0 fs time step.

**Path-Integral Formulation for Polarization**

The calculation of spontaneous polarization during continuous interlayer sliding presents a significant challenge. Conventional reference-structure-based methods often fail to properly treat periodic boundary conditions (PBC), resulting in unphysical discontinuities in the calculated polarization. To overcome this limitation and accurately capture the sliding ferroelectricity, we implemented a real-space path-integral formulation derived from the Modern Theory of Polarization. The continuous evolution of the macroscopic polarization component $P_\alpha(t)$ over time $t$ is defined as:

$$P_\alpha(t) = \int_{r(0)}^{r(t)} \frac{e}{\Omega} \sum_i Z^*_{i,\alpha\beta} \cdot dr_{i,\beta} \tag{1}$$

where $e$ is the elementary charge, $\Omega$ is the unit cell volume, and $Z^*_{i,\alpha\beta}$ represents the components of the BEC tensor for atom $i$.

For the discrete molecular dynamics trajectories, this path integral is computationally realized by accumulating the polarization increments at each simulation step $t$ up to the total steps $T$:

$$P_\alpha(T) = \sum_{t=1}^{T} \frac{e}{\Omega(t)} \sum_i \sum_\beta \frac{Z^*_{i,\alpha\beta}(t-1) + Z^*_{i,\alpha\beta}(t)}{2} \Delta r_{i,\beta}(t) \tag{2}$$

where $\Omega(t)$ is the dynamic cell volume. To strictly enforce PBC, the fractional coordinates $s_i(t)$ are utilized to determine the wrapped Cartesian displacement $\Delta r_i(t)$:

$$\Delta r_i(t) = [s_i(t) - s_i(t-1) - round(s_i(t) - s_i(t-1))]\bar{A}(t) \tag{3}$$

where $\bar{A}(t)$ denotes the average unit cell matrix between adjacent steps. Prior to the integration, an Acoustic Sum Rule (ASR) correction was systematically applied to the ML-predicted raw BEC tensors ($\tilde{Z}_i^* = Z_i^* - \frac{1}{N}\sum_i^N Z_i^*$) to guarantee strict charge neutrality.

**State-Constrained Time-Domain Background Extraction**

In the presence of an external out-of-plane electric field ($E_z$), the total calculated polarization ($P_z$) inevitably contains a linear dielectric background. To rigorously isolate the intrinsic spontaneous out-of-plane polarization from this dynamic dielectric response, we developed a state-constrained time-domain extraction protocol.

First, any accumulated integration drift over the total MD steps $N$ is linearly corrected:

$$P_z^{dc}(t_n) = P_z(t_n) - \frac{n}{N}[P_z(t_N) - P_z(t_0)] \tag{4}$$

where $P_z^{dc}(t_n)$ is the drift-corrected polarization at step $n$. Next, we construct a stacking determination function $S(t)$ that assigns +1 for the AB stacking state and -1 for the BA state based on the structural instantaneous configuration. The ideal spontaneous polarization is then approximated as $P_{ideal}(t_n) = P_s \cdot S(t_n)$, where $P_s$ is the average spontaneous polarization magnitude defined by $P_s = \frac{1}{2}(\overline{P_s^{AB}} - \overline{P_z^{BA}})$.

The raw background $B_{raw}(t_n) = P_z^{dc}(t_n) - P_{ideal}(t_n)$, which contains both low-frequency field-induced components and high-frequency thermal noise. To exclusively remove the field-induced background while preserving the physical thermal fluctuations, the trajectory is segmented according to the stacking state, and $B_{raw}$ is smoothed via a Gaussian convolution filter:

$$B_{smooth}(t) = \int B_{raw}(\tau)G(t-\tau;\sigma)d\tau \qquad (5)$$

Finally, the pure ferroelectric hysteresis loops are obtained by subtracting this smoothed dielectric background from the drift-corrected polarization.

**Conclusion**

In summary, we have developed a fully coupled machine-learning atomistic simulation framework integrating a vdW-inclusive fine-tuned MACE potential with a real-time EGCNN-based BEC predictor, enabling non-equilibrium MD simulations of electric-field-driven polarization switching in bilayer h-BN. The MACE model accurately reproduces experimental interlayer spacing, potential energy surface topology, and phonon dispersions, while the EGCNN achieves high BEC predictive precision across all structural phases.

Non-equilibrium MD simulations revealed that polarization switching proceeds via coherent single-domain rigid interlayer translation along symmetry-equivalent [$1\bar{1}0$] directions, completing within approximately 5 ps. The simulated coercive field of ±1.5 V/Å reflects the intrinsic switching limit of a defect-free lattice, consistent with experimental devices switching at lower fields via heterogeneous domain nucleation. The state-constrained Gaussian convolution background extraction successfully recovers clean, symmetric hysteresis loops consistent with experimental observations, with residual quantitative discrepancies attributed to non-equilibrium thermal contributions not fully separable within the current framework. The present work establishes a rigorous computational foundation for the predictive atomistic design of 2D sliding ferroelectric devices.

**Acknowledgements**

Y. Wang acknowledges support from the Program for Leading Graduate Schools (MERIT-WINGS). P. Chen would acknowledge the support of JST SPRING (Grant Number JPMJSP2108). This study was supported by the Ministry of Education, Culture, Sports, Science and Technology (MEXT) (Nos. 24H00042 and 26K01205).

**Data and Code Availability**

The data and code used for this article will be open after publication.

**Competing Interest**

The authors declare no competing interests.

# Ultrafast Sliding Ferroelectric Switching in Bilayer Hexagonal Boron Nitride Revealed by Deep Learning Molecular Dynamics

Yinan Wang[1,2], Poyen Chen[1,2], Teruyasu Mizoguchi[2]

3. Department of Materials Engineering, the University of Tokyo, Tokyo, Japan
4. Institute of Industrial Science, the University of Tokyo, Tokyo, Japan

**S1. Data Distribution**

To ensure balanced configurational coverage across the full structural phase space of bilayer h-BN, training data for both models were assembled from seven structural categories: AA′, AA′ sliding, AB, AB sliding, BA, moiré superlattices, and monolayer configurations. The MACE fine-tuning dataset comprises 2,808 structures sampled from AIMD trajectories and static sliding pathways, while the EGCNN training dataset comprises 4,045 structures with DFT-computed BEC tensors. The notably larger representation of moiré structures reflects the structural complexity and atomic diversity of the $\sum 7$, $\sum 13$, and $\sum 19$ commensurate superlattices. For both datasets, a stratified 8:2 train–test split was applied independently within each structural category to prevent dataset imbalance from biasing model generalization. The following Fig.S1 represents the distribution of the data used for fine-tunning MACE model and training the EGCNN model.

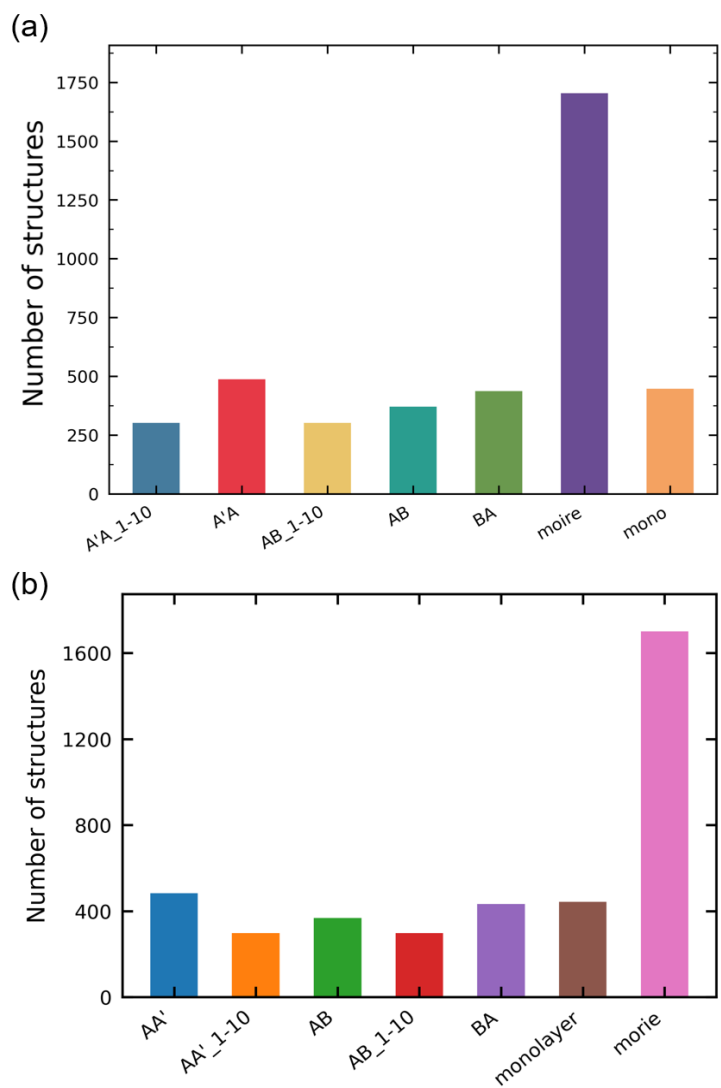

Fig.S1 Structural composition of the training datasets. (a) Distribution of the 2,808 structures used for fine-tuning the MACE machine learning potential, comprising seven structural subsets: monolayer h-BN, AA′, AA′ sliding (AA′ 1-10), AB, AB sliding (AB_1-10), BA, moiré superlattices (∑7 commensurabilities), and monolayer configurations. (b) Distribution of the 4,045 structures with DFT-computed BECs used for training the EGCNN model, comprising six structural subsets: monolayer h-BN, AA′, AA′ sliding (AA′ 1-10), AB, AB sliding (AB_1-10), BA, moiré superlattices (∑7, ∑13, ∑19 commensurabilities).

**S2. Fine-tuned MACE**

The predictive performance of the fine-tuned MACE model was evaluated on held-out test sets drawn from each structural subset. Parity plots comparing MACE-predicted total energies and atomic forces against DFT reference values are presented in Fig. S2. Across all structural categories, including equilibrium stackings (AA′, AB, BA), interlayer sliding pathways (AA′_1-10, AB_1-10), moiré superlattices, and monolayer configurations, the predicted values fall tightly along the perfect-prediction diagonal. Quantitative evaluation yields a training RMSE of 19.3 meV/atom for energies and 62.3 meV/Å for forces. The near-identical scatter observed across all structural subsets confirms that the fine-tuned model generalizes uniformly across the full configurational phase space without overfitting to any particular stacking type.

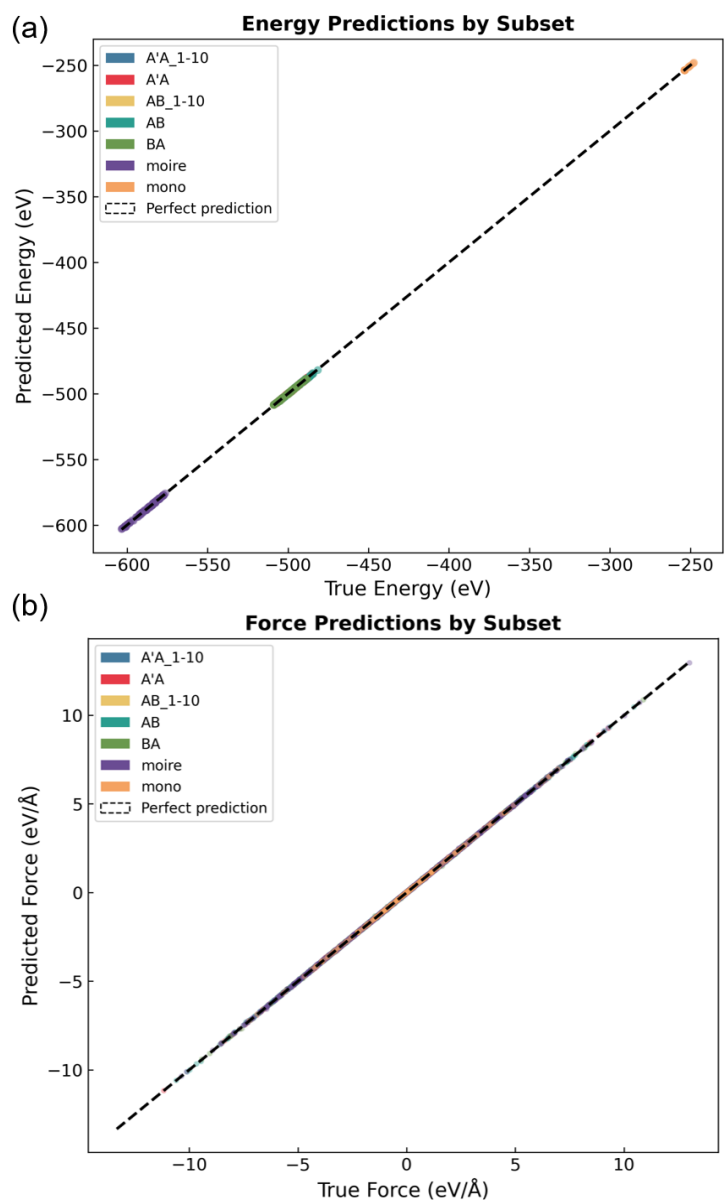

Fig.S2 Parity plots of the fine-tuned MACE machine learning potential evaluated on held-out test structures. (a) Predicted versus DFT total energies (eV) across all structural subsets. (b) Predicted versus DFT atomic force components (eV/Å) across all structural subsets. Data points are color-coded by structural category: monolayer (mono), AA′, AA′ sliding (AA′_1-10), AB, AB sliding (AB_1-10), AB, BA, and moiré superlattices. The dashed diagonal line indicates perfect prediction.

## S3. Atomic Trajectory Snapshots of Electric-Field-Driven Switching

To provide direct structural visualization of the electric-field-driven polarization switching dynamics, representative snapshots extracted from the non-equilibrium MD trajectory are presented in Fig. S3. The simulation was performed on a 20 Å supercell (256 atoms) at 50 K under a cyclic stepwise out-of-plane electric field swept along the sequence 0→−2→0→+2→0 V/Å.

At 0 ps, the system resides in the initial AB stacking configuration. The structure remains stable in this state until the applied field reaches the coercive threshold, at which point the first switching event occurs between 160 ps and 165 ps. The snapshots at 161–165 ps capture the transient structural evolution during this AB→BA transition, completing within approximately 5 ps. The concerted, simultaneous translation of both layers is clearly visible as a progressive lateral displacement of the atomic sublattices. Following the transition, the system stabilizes in the BA stacking state, which is retained through 560 ps as the field reverses polarity. The second switching event, corresponding to the BA→AB back-transition, proceeds analogously between 560 ps and 565 ps. At 820 ps, the system has returned to the AB stacking configuration, completing a full ferroelectric hysteresis cycle.

These snapshots confirm that the switching mechanism proceeds via coherent single-domain rigid sliding, with no evidence of partial domain formation or intermediate disordered states, consistent with the sharp, deterministic stacking transitions reported in Fig.3a of the main text.

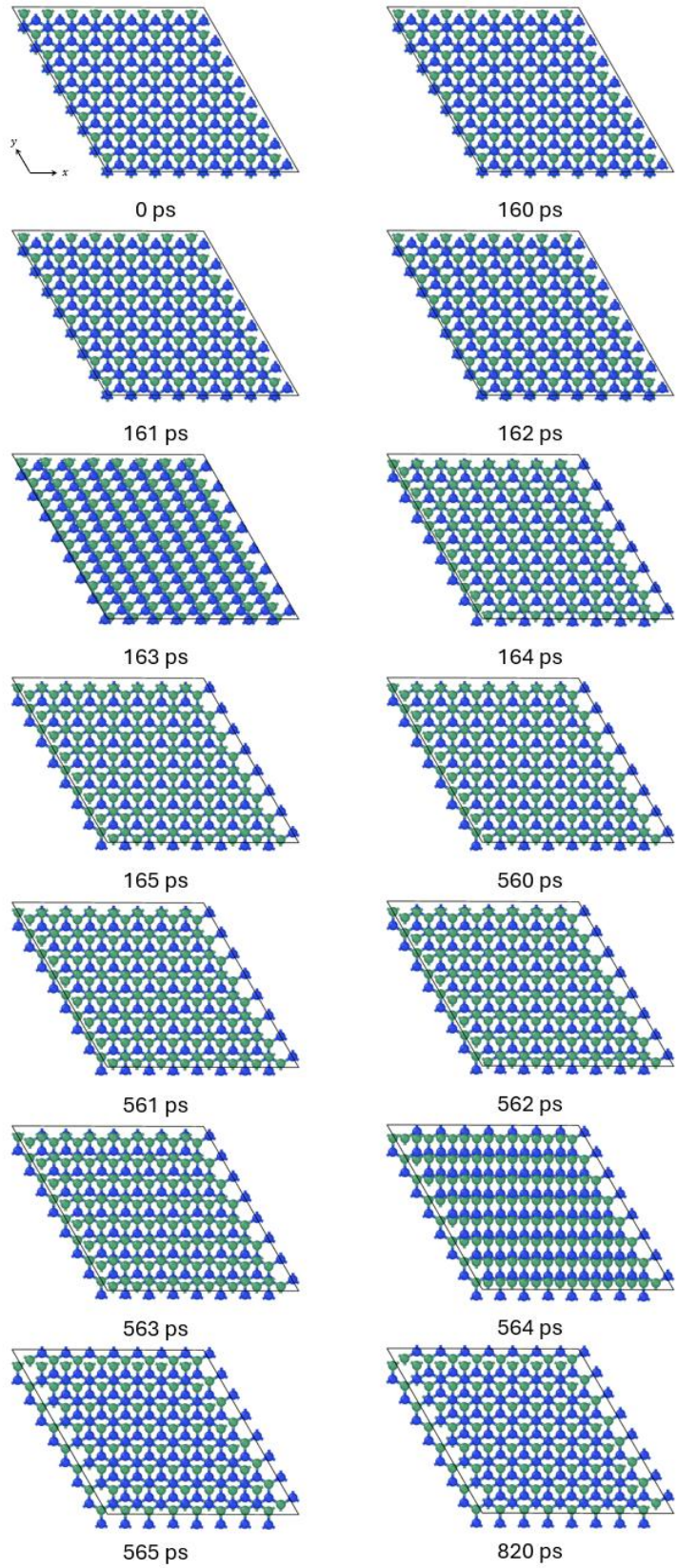

Fig.S3 Atomic trajectory snapshots of bilayer h-BN during electric-field-driven polarization switching, viewed along the out-of-plane direction. Green and blue spheres represent boron and nitrogen atoms, respectively. Atomic configurations were visualized using OVITO. Snapshots are extracted at selected time steps: 0 ps (initial AB stacking), 160 ps (pre-switching), 161–165 ps (AB→BA transition), 560 ps (pre-switching), 561–565 ps (BA→AB back-transition), and 820 ps (recovered AB stacking). The coherent lateral displacement of the atomic sublattices across the transition window confirms the single-domain rigid sliding mechanism.

To structurally quantify the interlayer sliding dynamics at each field step, the root-mean-square displacement (RMSD) of atomic positions was computed separately for the in-plane (xy) and out-of-plane (z) components throughout the full cyclic field sweep (Fig. S4). At each field amplitude, the system was evolved for 20 ps, and the RMSD was evaluated relative to the initial configuration of that field step.

Throughout the majority of the sweep, the in-plane RMSD (red) remains at a stable, elevated plateau (~0.8 Å) corresponding to the lateral offset between AB and BA sublattice registries, while the out-of-plane RMSD (blue) remains near zero, confirming that the interlayer spacing is preserved and no out-of-plane structural distortion occurs. At the coercive field steps ($\varepsilon_z = -1.6$ V/Å and $\varepsilon_z = +1.6$ V/Å), the in-plane RMSD exhibits a sharp, transient spike followed by convergence to a new plateau value, directly capturing the moment of stacking inversion. The z-component remains flat throughout all switching events, confirming that the transition involves purely in-plane rigid translation with no buckling or interlayer separation. This structural evidence is fully consistent with the coherent single-domain sliding mechanism described in the main text.

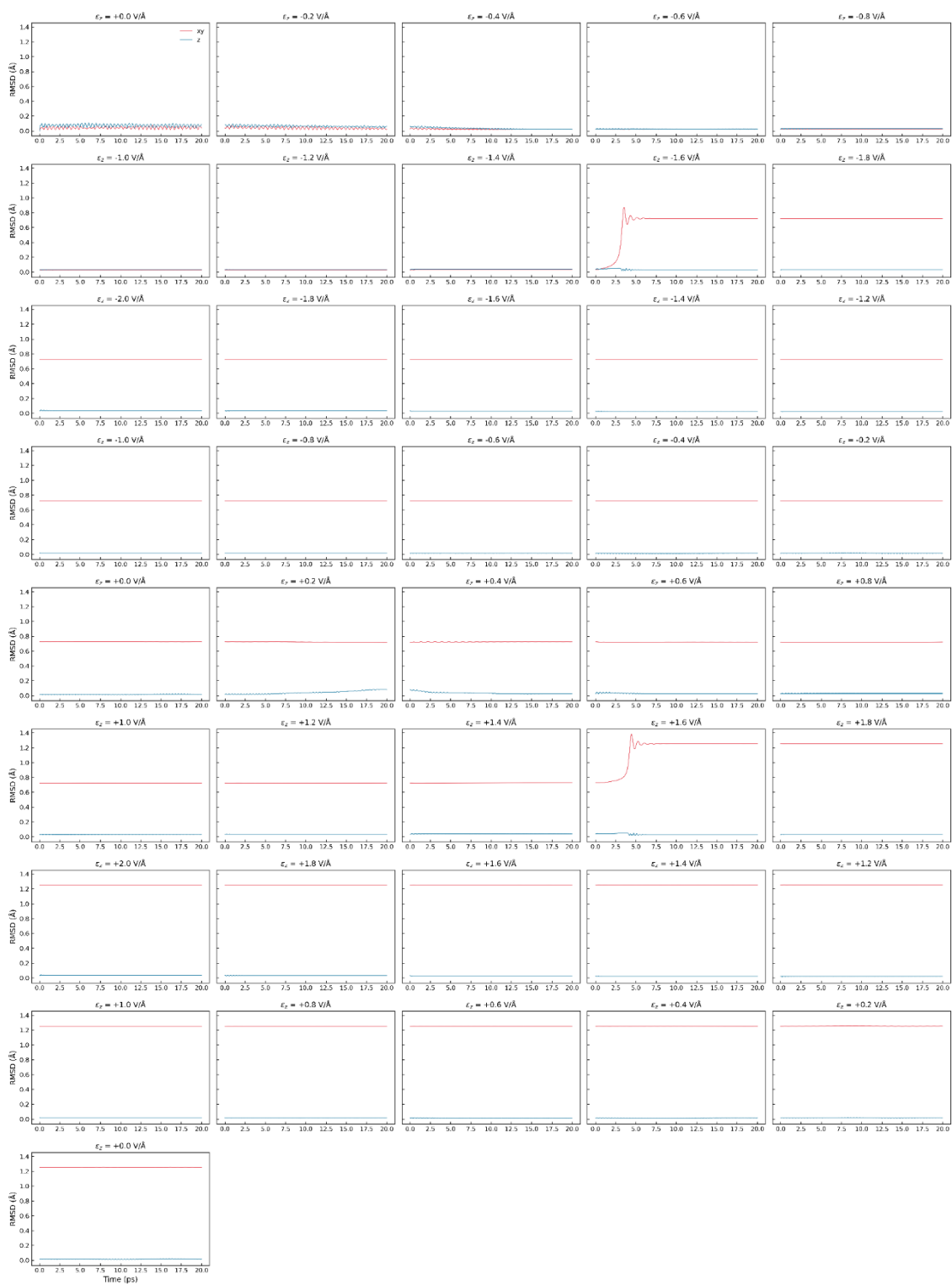

Fig.S4 Per-field-step RMSD analysis of the full cyclic electric-field-driven MD trajectory. Each panel shows the time evolution of the in-plane (xy, red) and out-of-plane (z, blue) RMSD components over 20 ps at the indicated field amplitude $\varepsilon_z$, sequentially following the sweep

0→−2→0→+2→0 V/Å. Sharp transient spikes in the xy RMSD at $\varepsilon_z = -1.6$ V/Å and +1.6 V/Å mark the AB↔BA stacking inversion events, while the z-component remains negligible throughout, confirming that polarization switching proceeds via purely in-plane coherent rigid sliding without structural disordering.

## S4. Temperature Dependence of Electric-Field-Driven Switching and Extracted Spontaneous Polarization

To systematically investigate the influence of temperature on the electric-field-driven switching behavior, non-equilibrium MD simulations were performed at 100 K, 200 K, and 300 K under identical cyclic field sweep conditions (0→−2→0→+2→0 V/Å) in a 20 Å supercell. For each temperature, the stacking state evolution (left column), the raw cumulative $P_z$ trajectory (middle column), and the background-subtracted ferroelectric hysteresis loop (right column) are presented in Fig. S5.

Across all three temperatures, the stacking state panels (left column) confirm sharp, deterministic AB↔BA switching events occurring at a consistent coercive field of approximately ±1.5 V/Å, with no discernible temperature dependence in the transition threshold. This athermal character of the coercive field is consistent with the electromechanical in-plane force transduction mechanism discussed in the main text.

The raw $P_z$ trajectories (middle column) exhibit the characteristic rhombus-shaped profile arising from the superposition of the intrinsic spontaneous polarization and the cumulative field-induced dielectric background, as described in the main text. The magnitude of the raw signal increases systematically with temperature, reflecting enhanced thermal atomic displacements that amplify the cumulative path-integral background at elevated temperatures.

Following the state-constrained Gaussian convolution background subtraction procedure, the purified hysteresis loops (right column) recover flat saturation plateaus with abrupt switching transitions. The extracted remanent spontaneous polarization magnitudes are $P_S$= 2.84 μC/cm² at 100 K, $P_S$= 7.33 μC/cm² at 200 K, and $P_S$= 12.13 μC/cm² at 300 K. The systematic increase of the extracted $P_S$ with temperature is attributed to residual non-equilibrium thermal contributions that are not fully separable within the current signal processing framework, as acknowledged in the main text. The intrinsic zero-temperature spontaneous polarization of AB-stacked bilayer h-BN, as derived from static path-integral calculations using ASR-corrected EGCNN BECs (Fig.2b, main text), provides the physically meaningful reference value independent of these dynamic thermal artifacts.

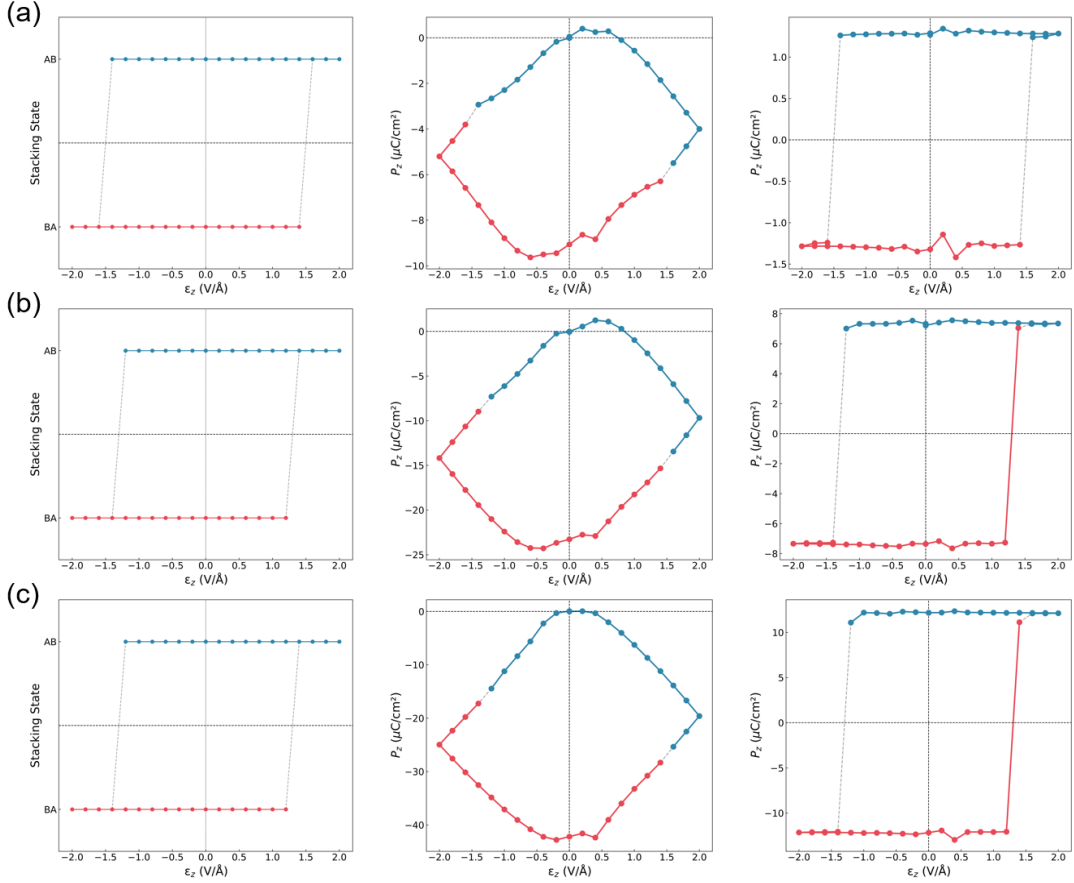

Fig. S5 Temperature dependence of electric-field-driven switching dynamics in bilayer h-BN. Results are shown for (a) 100 K, (b) 200 K, and (c) 300 K, each under a cyclic field sweep of 0→−2→0→+2→0 V/Å in a 20 Å supercell. Left column: temporal evolution of the stacking state (AB, blue; BA, red) as a function of applied field $\varepsilon_z$. Middle column: corresponding raw cumulative $P_z$ as a function of $\varepsilon_z$. Right column: purified ferroelectric hysteresis loops obtained after state-constrained Gaussian convolution background subtraction. The extracted remanent spontaneous polarization magnitudes are $P_s$= 2.84 μC/cm² (100 K), 7.33 μC/cm² (200 K), and 12.13 μC/cm² (300 K).

## S5. Thermodynamic Estimate of the Intrinsic Coercive Field

The thermodynamic coercive field $\varepsilon_c$ can be estimated by minimizing the electric enthalpy of the system. Under an applied out-of-plane electric field $\varepsilon_z$, the electric enthalpy $H$ per unit cell is defined as:

$$H = E - P\varepsilon \qquad (S1)$$

where $E$ is the potential energy, $P$ is the macroscopic polarization, and $\varepsilon$ is the applied electric field. At the coercive field, the electric enthalpy difference between the stable stacking state and the saddle point is balanced by the electrostatic energy gained from the applied field. This condition yields a simple thermodynamic estimate for $\varepsilon_c$:

$$\varepsilon_c = \frac{\Delta E_{SP}}{A_{xy} P_s c} \qquad (S2)$$

where $\Delta E_{SP}$ is the energy barrier at the saddle point extracted from the CI-NEB calculations (Fig. 1d), $A_{xy}$ is the in-plane cross-sectional area of the supercell, $P_s$ is the intrinsic spontaneous polarization, and $c$ is the out-of-plane lattice parameter. Substituting the DFT-computed values yields an intrinsic coercive field of approximately 39.8 V/Å, which drastically overestimates the simulated $\varepsilon_c$ of ±1.5 V/Å. As discussed in the main text, this overestimation demonstrates that the switching is not driven solely by the direct coupling between $\varepsilon_z$ and $P_z$, but is instead dominated by the off-diagonal BEC components $Z^*_{zx}$ and $Z^*_{zy}$, which transduce the out-of-plane field into in-plane sliding forces.

## S6. Structural Validation of the Free-Standing Bilayer Approximation

Prior theoretical studies have reported that in certain device configurations, nitrogen atoms in h-BN can form covalent bonds with underlying substrate materials such as SiC, raising the question of whether interfacial bonding might perturb the electronic structure and dynamics of the ferroelectrically active layer.[50] To address this concern and to validate the free-standing bilayer approximation adopted in our MD simulations, we constructed a heterogeneous supercell replicating the experimental device geometry reported by Yasuda et al.[1] Specifically, the model consists of a central AB-stacked bilayer h-BN flanked on each side by a three-layer AA′-stacked h-BN slab, with a single graphene layer inserted between the AB bilayer and each AA′ slab, yielding a symmetric stack of the form AA′(3L)/graphene/AB(2L)/graphene/AA′(3L). For each interface, the interlayer stacking registry was set to the thermodynamically stable configuration.[51] Static DFT calculations were subsequently performed on this full heterostructure.

The calculated interlayer spacings between the AB bilayer and the adjacent graphene layers, as well as between the graphene layers and the outer AA′ slabs, all remain above 3 Å, consistent with the interlayer distances obtained for isolated bilayer h-BN and graphene systems with an explicit vacuum layer. No evidence of covalent bond formation is found across any of these interfaces. This result demonstrates that the AB bilayer resides in an effectively free-standing electrostatic environment, coupled to its surroundings exclusively through van der Waals interactions. Accordingly, the approximation of simulating the ferroelectrically active AB bilayer h-BN with vacuum boundary conditions in our non-equilibrium MD framework is physically justified, and the absence of interfacial covalent bonding ensures that the polarization switching dynamics reported in the main text are not perturbed by substrate or capping layer effects.

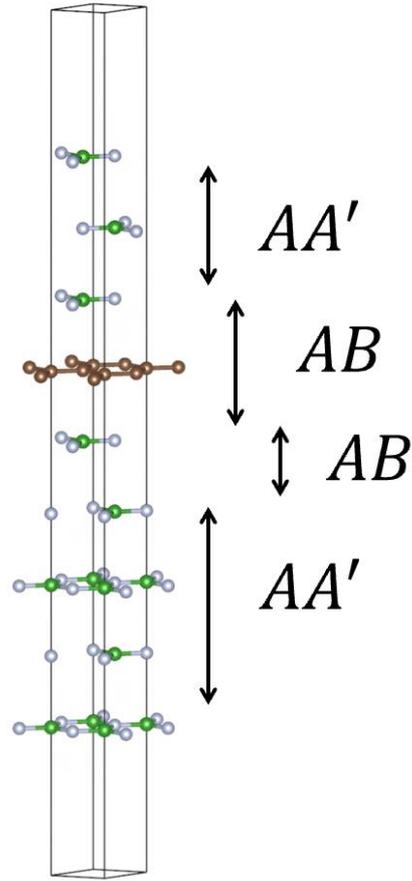

Fig.S6 Side view of the heterogeneous supercell constructed to validate the free-standing bilayer approximation. The structure consists of a central AB-stacked bilayer h-BN sandwiched between a single graphene layer (brown) and a three-layer AA′-stacked h-BN slab on each side, forming a symmetric AA′(3L)/graphene/AB(2L)/graphene/AA′(3L) stack. Green and gray spheres represent boron and nitrogen atoms, respectively. The interlayer stacking registry at each interface is set to the thermodynamically stable configuration. Vacuum layers are applied above and below the full heterostructure to eliminate interactions between periodic images along the out-of-plane direction.

S8. Supercell size dependance

It is imperative to establish the thermodynamic boundary conditions under which the ferroelectric states remain stable against thermal fluctuations in the absence of an external electric field. We performed zero-field MD simulations to investigate the retention of the initial AB stacking state across various temperatures and supercell sizes (Fig.S7). The results demonstrate a pronounced finite-size and temperature dependence, rooted in the collective nature of the sliding mechanism. For small supercells (e.g., 10 Å) or at elevated temperatures exceeding 300 K, local thermal fluctuations can easily percolate across the limited periodic boundaries. This provides sufficient thermal kinetic energy to overcome the shallow SP barrier, causing the AB stacking ratio to plummet as the layers spontaneously and randomly slide.

Conversely, in larger supercells ($\geq 20$ Å), the sliding mechanism requires the coherent collective movement of a substantially larger number of atoms. Because the absolute total kinetic barrier scales extensively with the interfacial area, spontaneous thermal transitions are effectively statistically suppressed at temperatures up to 300 K. These simulation results not only justify better MD parameters to isolate deterministic electric-field-driven switching from stochastic thermal noise, but also provide an atomistic explanation for macroscopic experiments. By extrapolating this finite-size stabilization to the micrometer scale of realistic device channels, our findings align with the experimental robust retention of sliding ferroelectricity in h-BN devices at room temperature.

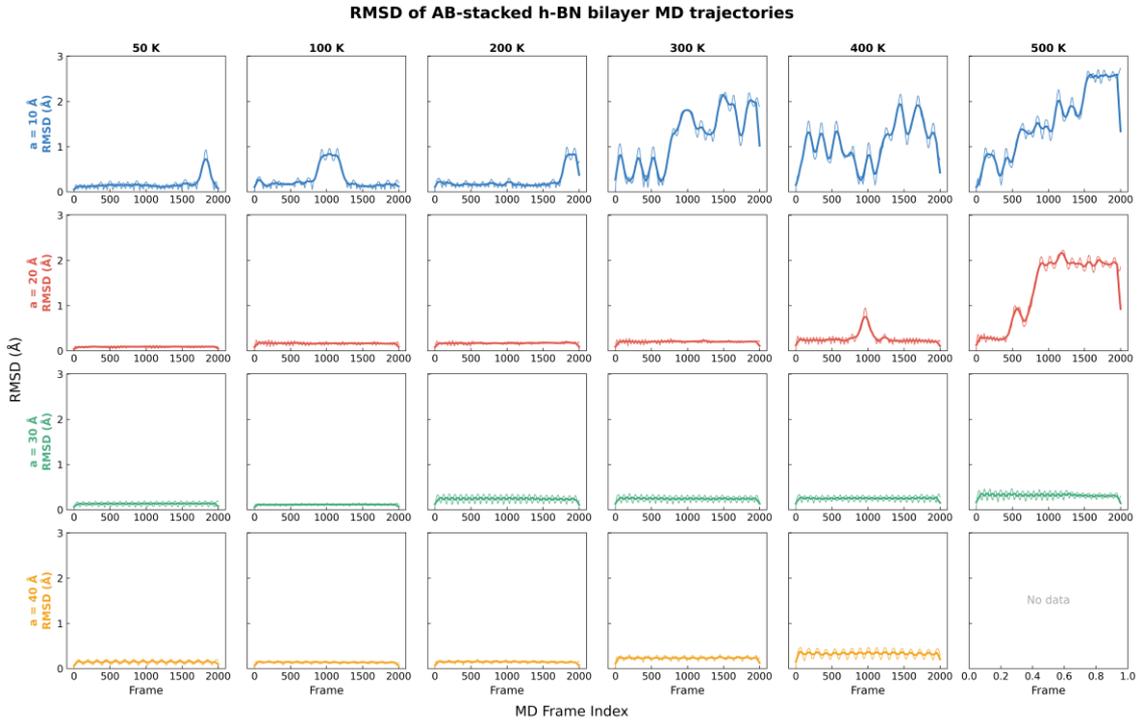

Fig.S7 Thermodynamic stability of the AB stacking state under zero-field NVT molecular dynamics as a function of supercell size and temperature. Each panel shows the time evolution of the total atomic RMSD computed relative to the initial AB-stacked reference configuration across 2,000 frames (~20 ps) at the indicated temperature (50–500 K) and in-plane lattice parameter $a$ (10, 20, 30, and 40 Å, corresponding to supercells of 32, 128, 288, and 512 atoms, respectively).